\documentclass[12pt,a4paper,final]{iopart}

\usepackage{iopams}
\usepackage{graphicx}

\usepackage{amssymb,amsthm}
\usepackage{amscd}	
\usepackage{enumerate}
\usepackage{framed}
\usepackage{subfigure}
\usepackage{xcolor}
\usepackage{color}
\usepackage{cases}
\usepackage[margin=1cm]{caption}

\newcommand{\bra}[1]{\mbox{$\left\langle #1 \right|$}}
\newcommand{\ket}[1]{\mbox{$\left| #1 \right\rangle$}}

\newcommand{\eqref}[1]{(\ref{#1})}

\eqnobysec

\begin{document}
\title{Reliable and robust entanglement witness}


\author{Xiao Yuan$^{*}$, Quanxin Mei, Shan Zhou, Xiongfeng Ma$^{\dag}$}
\address{Center for Quantum Information, Institute for Interdisciplinary Information Sciences, Tsinghua University, Beijing 100084, China}
\ead{$^*$yuanxiao12@mails.tsinghua.edu.cn, $^\dag$xma@tsinghua.edu.cn}

\begin{abstract}
Entanglement, a critical resource for quantum information processing, needs to be witnessed in many practical scenarios. Theoretically, witnessing entanglement is by measuring a special Hermitian observable, called entanglement witness (EW), which has non-negative expected outcomes for all separable states but can have negative expectations for certain entangled states. In practice, an EW implementation may suffer from two problems. The first one is \emph{reliability}. Due to unreliable realization devices, a separable state could be falsely identified as an entangled one. The second problem relates to \emph{robustness}. A witness may be suboptimal for a target state and fail to identify its entanglement. To overcome the reliability problem, we employ a recently proposed measurement-device-independent entanglement witness scheme, in which the correctness of the conclusion is independent of the implemented measurement devices. In order to overcome the robustness problem, we optimize the EW to draw a better conclusion given certain experimental data. With the proposed EW scheme, where only data postprocessing needs to be modified comparing to the original measurement-device-independent scheme, one can efficiently take advantage of the measurement results to maximally draw reliable conclusions.
\end{abstract}

\vspace{2pc}
\maketitle

\section{Introduction}
Since the inception of quantum theory, entanglement has been recognized as one of the most distinctive quantum features. In a way, Einstein, Podolsky, and Rosen proposed a paradox \cite{Einstein35} on entanglement, which was motivated to argue against the quantum theory, however turned out to be an effective experimental (Bell) test \cite{bell} for ruling out classical theories. In the development of the quantum information field, entanglement becomes an essential resource for varieties of tasks \cite{Horodecki09}. Many quantum advantages can be revealed if there exists entanglement. Witnessing the existence of entanglement is thus an important and necessary step for quantum information processing. For instance, in quantum key distribution (QKD) \cite{bb84,Ekert91}, secret keys are ensured crucially by showing that entanglement can be preserved after the quantum channel \cite{Lutkenhaus04}. In quantum computing, witnessing the existence of entanglement is an important benchmark for the following experiment \cite{tura2014detecting}.

\begin{figure}[bht]
\centering
\resizebox{8cm}{!}{\includegraphics[scale=1]{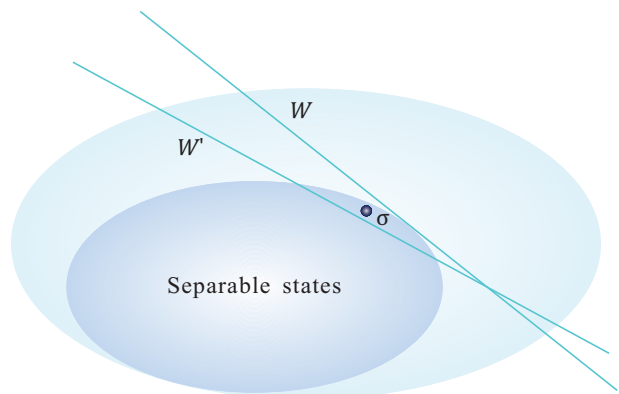}}
\caption{Entanglement witness and the reliability problem. }\label{fig:Witnesscw}
\end{figure}

In theory, as shown in Fig.~\ref{fig:Witnesscw}, entanglement can be witnessed by measuring a Hermitian observable $W$, whose output expectation for any separable state $\sigma$ is non-negative,
\begin{equation}\label{OMDIEW:EWsep}
\mathrm{Tr}(W\sigma)\ge0,
\end{equation}
but can be negative for certain entangled state $\rho$,
\begin{equation}\label{OMDIEW:EWent}
\mathrm{Tr}(W\rho)<0.
\end{equation}
In this case, we call $W$ an entanglement witness (EW) for $\rho$. In general, $W$ can be obtained by a linear combination of product observables, which can be measured locally on the subsystems \cite{guhne2009}. 

In reality, EW implementation may suffer from two problems. The first one is \emph{reliability}. That is, one might conclude unreliable results due to imperfect experimental devices. In this case, the validity of the EW result depends on how faithful one can implement the measurements according to the witness $W$. If the realization devices are not well calibrated, the practically implemented observable $W'$ may deviate from the original theoretical design $W$, see Fig.~\ref{fig:Witnesscw} as an example, which can even be not a witness. That is, there may exist some separable states $\sigma$, such that $\mathrm{Tr}[\sigma W'] < 0 \le \mathrm{Tr}[\sigma W]$. Practically, by exploiting device imperfections, an attack has been experimentally implemented for an entanglement witness procedure \cite{Yuan14}. In cryptographic applications, such problem is regarded as a loophole, where one mistakes separable states to be entangled ones. For instance, in QKD, this would indicate that an adversary successfully convinces the users Alice and Bob to share keys which they think are secure but are eavesdropped.  Such problem is solved by the measurement-device-independent QKD scheme \cite{Lo2012MDI}, inspired by the time-reversed entanglement-based scheme \cite{Biham:1996:Quantum,Inamori:TimeReverseEPR:2002,Stefano:MDIQKD:2012}. Branciard et al.~applied a similar idea to EW and proposed the measurement-device-independent entanglement witness (MDIEW) scheme \cite{Branciard13}, in which entanglement can be witnessed without assuming the realization devices. The MDIEW scheme is based on an important discovery that any entangled state can be witnessed in a nonlocal game with quantum inputs \cite{Buscemi12}. In the MDIEW scheme, it is shown that an arbitrary conventional EW can be converted to be an MDIEW, which has been experimentally tested \cite{Yuan14}.


The second problem lies on the \emph{robustness} of EW implementation. Since each (linear) EW can only identify certain regime of entangled states, a given EW is likely to be ineffective to detect entanglement existing in an unknown quantum state. While a failure of detecting entanglement is theoretically acceptable, in practice, such failure may cause experiment to be highly inefficient. In fact, a conventional EW can only be designed optimal when the quantum state has been well calibrated, which, on the other hand, generally requires to run quantum state tomography.
Practically, when the prepared state can be well modeled, one can indeed choose the optimal EW to detect its entanglement. Since a full tomography requires exponential resources regarding to the number of parties, EW plays as an important role for detecting well modeled entanglement, which would generally fail for an arbitrary unknown state.
In a way, this problem becomes more serious in the MDIEW scenario, where the measurement devices are assumed to be uncharacterized and even untrusted. In this case, the implemented witness, which may although be designed optimal at the first place, can become a bad one which merely detects no entanglement. However, the observed experimental data may still have enough information for detecting entanglement.
Therefore, the key problem we are facing here is that given a set of observed experimental data, what is the best entanglement detection capability one can achieve. That is, we want to maximize the detectable entangled states with a fixed experimental setup.

\begin{figure*}[bht]
\centering
\resizebox{14cm}{!}{\includegraphics[scale=1]{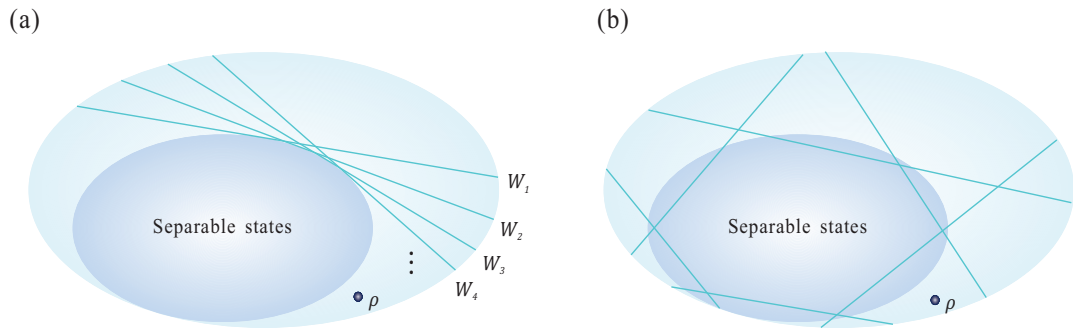}}
\caption{Optimization of entanglement witnesses. (a) To get the optimal witness of an unknown entangled state $\rho$, one has to run over all possible witnesses. Intuitively, this is done by scanning over all witnesses that are \emph{tangent} to the set of separable states. (b) The optimization can be efficiently done if certain failure probability can be tolerated.}\label{fig:witness2}
\end{figure*}

In detecting quantum nonlocality, a similar problem is to find the optimal Bell inequality for the observed correlation, which can be solved efficiently with linear programming \cite{boyd2004convex}.
Regarding to our problem, we essentially need to optimize over all entanglement witness to draw the best conclusion of entanglement with the same experiment data, as shown in Fig.~\ref{fig:witness2}(a). As the set of separable states is not a polytope, this problem cannot be solved by linear programming. Generally speaking, it is proved that the problem of accurately finding such an optimal witness is NP-hard \cite{ben1998robust}. However, if certain failure probability is tolerable, we show in this work that this problem can be efficiently solved. That is, if we admit a probability less than $\epsilon$ to detect a separable state to be entangled, we show that the optimal entanglement witness can be efficiently found. As the optimization step can be effectively conducted as post-processing, our scheme does not pose extra burdens to experiments compared to the original MDIEW scheme. In this case, our result can be directly applied in practice.




The rest of the paper is organized as follows. In Section \ref{Sec:MDIEW}, we review the MDIEW scheme, which solves the reliability problem. Then, we introduce our robust MDIEW scheme in Section \ref{Sec:OMDIEW} and give an explicit example in Section \ref{Sec:Example}. In Section \ref{Sec:Discussion}, we conclude our result and discuss practical applications. We mainly focus our discussion on the bipartite scenario. While, our result can be naturally generalized to multipartite cases.

\section{Reliable entanglement witness}\label{Sec:MDIEW}
\subsection{Nonlocal game}
\begin{figure}[bht]
\centering
\resizebox{8cm}{!}{\includegraphics[scale=1]{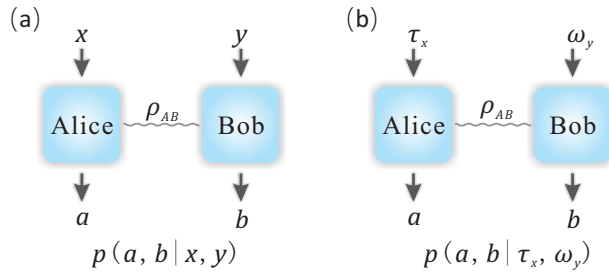}}
\caption{Bipartite nonlocal game with classical and quantum inputs. (a) Nonlocal game with classical inputs. Based on the classical inputs $x$ and $y$, Alice and Bob perform local measurement on the pre-shared entangled state $\rho_{AB}$, and get classical outputs $a$ and $b$, respectively. A linear combination of the probability distribution $p(a,b|x,y)$ defines a Bell inequality as shown in Eq.~\eqref{eq:Bell}. 
(b) Nonlocal game with quantum inputs. The quantum inputs of Alice and Bob are respectively $\tau_x$ and $\omega_y$. It is shown \cite{Buscemi12} that any entangled quantum states can be witnessed with a certain nonlocal game with quantum inputs. Equivalently, if we consider that Alice and Bob each prepares an ancillary state and a third party Eve performs the measurement, this setup also corresponds to the case of MDIEW. }\label{fig:nonlocal}
\end{figure}

Before reviewing the MDIEW scheme, we first discuss about nonlocal games with classical and quantum inputs as shown in Fig.~\ref{fig:nonlocal}. In a classical nonlocal game, classical random inputs $x$ and $y$ are given to two spacelikely separated users Alice and Bob, who perform measurement on pre-shared entangled state $\rho_{AB}$ and output $a$ and $b$, respectively. According to the probability distribution $p(a,b|x,y)$, a Bell inequality can be defined by
\begin{equation}\label{eq:Bell}
I = \sum_{a,b,x,y} \beta_{a,b}^{x,y}{p}(a,b|x,y) \leq I_C,
\end{equation}
where $I_C$ is a bound for all separable state $\sigma_{AB}$. A violation of the inequality can be considered as a witness for entanglement. As the Bell test does not assume measurement detail, witnessing entanglement by Bell test is device independent. However, as the conclusion is so strong such that the implementation is self-testing, not all entangled states can be witnessed in such a way \cite{Werner89, Barrett02}. Furthermore, the requirement of a faithful Bell test is very high, which makes such a witnesses impractical. For instance, the minimum efficiency required is $2/3$ for all Bell tests with binary inputs and outputs \cite{Massar03, Wilms08}.
On the other hand, if we can trust the measurement, a Bell test essentially becomes an EW. Although such method is able to detect all entangled state and is easy to realize, this scheme is not measurement-device-imperfection-tolerant.


In the seminal work \cite{Buscemi12}, Buscemi introduces the concepts of nonlocal games with quantum inputs. Denote the inputs of Alice and Bob by $\omega_x$ and $\tau_y$, then an inequality similar to Bell inequality can be defined by
\begin{equation}\label{eq:QuantumBell}
J = \sum_{a,b,x,y} \beta_{a,b}^{x,y}{p}(a,b|\omega_x,\tau_y) \leq J_C,
\end{equation}
where $J_C$ is also the bound for all separable state $\rho_{AB}$. As the quantum inputs can be indistinguishable, it is proved that all entangled states can violate a certain inequality \cite{Buscemi12}. If we consider the input states are faithfully prepared by Alice and Bob, then such nonlocal game with quantum inputs can be considered as an MDIEW \cite{Branciard13}. Moreover, as shown below, there is no detection efficiency limit for such a test.

\subsection{MDIEW}
The nonlocal game presented in Ref.~\cite{Buscemi12} can be considered as a reliable entanglement witness method, which does not witness separable state as entangled with arbitrary implemented measurement. This nonlocal game is thus an MDIEW, i.e., $J\ge 0$ for all separable states and $J$ can be negative if Alice and Bob share entangled state. Furthermore, the statement that $J\ge 0$ for all separable states is independent of the implementation of the measurement. In Ref.~\cite{Branciard13}, the authors put this statement into more concrete and practical framework. They show that, for an arbitrary conventional EW, there is a corresponded MDIEW. Below, we will quickly show how to derive MDIEWs from conventional EWs.



Focus on the bipartite scenario with Hilbert space $\mathcal{H}_A\otimes \mathcal{H}_B$, with dimensions $\mathrm{dim}\mathcal{H}_A=d_A$ and $\mathrm{dim}\mathcal{H}_B=d_B$. For a bipartite entangled state $\rho_{AB}$ defined on $\mathcal{H}_A\otimes \mathcal{H}_B$, we can always find a conventional entanglement witness $W$ such that $\mathrm{Tr}[W\rho_{AB}] < 0$ and $\mathrm{Tr}[W\sigma_{AB}]\ge 0$ for any separable state $\sigma_{AB}$. Suppose $\{\omega_x\}$ and $\{\tau_y\}$ to be two bases for Hermitian operators on $\mathcal{H}_A$ and  $\mathcal{H}_B$, respectively. Thus, we can decompose $W$ on the basis $\{\omega_x\otimes\tau_y\}$ by
\begin{equation}\label{eq: decomposition}
	W = \sum_{x,y} \beta^{x,y}\omega_x^{\mathrm{T}} \otimes \tau_y^{\mathrm{T}},
\end{equation}
where $\beta^{x,y}$ are real coefficients and the transpose is for later convenience. Notice that, owing to the completeness of the set of density matrices, we further require $\{\omega_x\}$ and $\{\tau_y\}$ to be density matrices. In addition, the decomposition of Hermitian operators is not unique which varies with different $\{\omega_x\}$ and $\{\tau_y\}$.

With a conventional EW decomposed in Eq.~\eqref{eq: decomposition}, an MDIEW can be obtained by
\begin{equation}\label{eq: simplifiedBell}
	J = \sum_{x,y} \beta^{x,y}_{1,1} {p}(1,1|\omega_x,\tau_y)
\end{equation}
where $\beta^{x,y}_{1,1} =  \beta^{x,y}$ and ${p}(1,1|\omega_x,\tau_y)$ is the probability of outputting $(a=1,b=1)$ with input states $(\omega_x,\tau_y)$. In the MDIEW design, Alice (Bob) performs Bell state measurement on $\rho_A$ ($\rho_B$) and $\omega_x$ ($\tau_y$). The probability distribution ${p}(1,1|\omega_x,\tau_y)$ is thus obtained by the probability of projecting onto the maximally entangled state $\ket{\Phi_{AA}^+} =1/\sqrt{d_A}\sum_i \ket{ii}$ and $\ket{\Phi_{BB}^+} = 1/\sqrt{d_B} \sum_j \ket{jj}$.

As shown in Ref.~\cite{Branciard13} and also \ref{App:proof}, $J$ is linearly proportional to the conventional witness with ideal measurement,
\begin{equation}\label{}
  J = \mathrm{Tr}[W\rho_{AB}] / \sqrt{d_A d_B}.
\end{equation}
Thus, $J$ defined in Eq.~\eqref{eq: simplifiedBell} witnesses entanglement. Furthermore, it can be proved that such a witness is independent of the measurement devices. That is, even if the measurement devices are imperfect, $J$ is always non-negative for all separable states and hence no separable state will be mistakenly witnessed to be entangled. We refer to Ref.~\cite{Branciard13} and also \ref{App:proof} for a rigourous proof.

Theoretically, the MDIEW scheme prevents identifying separable states to be entangled. Such a reliable MDIEW has been experimentally demonstrated lately \cite{Yuan14}. In practice, however, such a scheme can be inefficient, meaning that it witnesses very few entangled states despite that the observed data could actually provide more information. This is because, in the MDIEW procedure, one first chooses a conventional EW and realize in an MDI way. The conventional EW is chosen based on an empirical estimation of the to-be-witnessed state, thus it may not be able to witness the state for an ill estimation. Furthermore, even if the conventional EW is optimal at the first place, the measurement imperfection will make it sub-optimal in practice. Especially, when the input states $\{\omega_x\otimes\tau_y\}$ is complete, a specific witness may not be able to detect entanglement. With complete information, a natural question is whether we can obtain maximal information about entanglement, i.e., get the optimal estimation of MDIEW.

\section{Robust MDIEW}\label{Sec:OMDIEW}
Now, we present a method to optimize the MDIEW given a fixed observed experiment data $p(1,1|\omega_x,\tau_y)$. Before digging into the details, we compare the problem to a similar one in nonlocality. In the nonlocality scenario, a Bell inequality is used as a witness for quantumness, see Eq.~\eqref{eq:Bell}. In practice, the Bell inequality may not be optimal for the observed probability distribution $p(a,b|x,y)$. As the probability distribution of classical correlation forms a polytope, one can run a linear programming to get an optimal Bell inequality for $p(a,b|x,y)$. While, in our case, the probability distribution $p(1,1|\omega_x,\tau_y)$ with separable states is only a convex set but no-longer a polytope. Thus, our problem cannot be solved directly with linear programming.

\subsection{Problem formulation}
Let us start with formulating the optimization problem. Informally, our problem can be described as follows,
\begin{flushleft}
\emph{Problem (informal): find an optimal witness for the observed probability distribution $p(1,1|\omega_x,\tau_y)$.}
\end{flushleft}
According to Eq.\eqref{eq: simplifiedBell}, the witness value is defined by a linear combination of $p(1,1|\omega_x,\tau_y)$ with coefficient $\beta^{x,y}$. To witness entanglement, the coefficient $\beta^{x,y}$ must lead to a witness as defined in Eq.~\eqref{eq: decomposition}. In addition, as we can always assign $2\beta^{x,y}$ to double a violation, we require a trace normalization of the witness $W$ by
\begin{equation}\label{}
  \mathrm{Tr}[W] = 1.
\end{equation}
Therefore, the problem can be expressed as
\begin{flushleft}
\emph{Problem (formal): For a given probability distribution ${p}(1,1|\omega_x,\tau_y)$, minimize
\begin{equation}\label{eq:Jqminimum}
J(\beta^{x,y}) = \sum_{x,y} \beta^{x,y}{p}(1,1|\omega_x,\tau_y)
\end{equation}
over all $\beta^{x,y}$ satisfying
\begin{equation}\label{eq:constraintsss}
\sum_{x,y}\beta^{x,y}\mathrm{Tr}\left[\sigma_{AB}(\omega_x^{\mathrm{T}}\otimes\tau_y^{\mathrm{T}})\right]\ge0,
\end{equation}
for any separable state $\sigma_{AB}$ and
\begin{equation}\label{}
  \mathrm{Tr}\left[\sum_{x,y} \beta^{x,y}\omega_x^{\mathrm{T}} \otimes \tau_y^{\mathrm{T}}\right] = 1.
\end{equation}
}
\end{flushleft}


Contrary to the optimization of Bell inequality, we can see that this problem is much more complex. When the measurements are implemented faithfully, it is easy to verify that $p(1,1|\omega_x,\tau_y) = \mathrm{Tr}[(\omega_x\otimes\tau_y)\rho_{AB}]/\sqrt{d_Ad_B}$, where $\rho_{AB}$ is the state measured. Therefore, finding the optimal $\beta^{x,y}$ is equivalent to find the optimal entanglement witness $W = \sum_{x,y}\beta^{xy}\omega_x^{\mathrm{T}}\otimes\tau_y^{\mathrm{T}}$ for state $\rho_{AB}$. A possible solution to this problem is to try all entanglement witnesses to find the optimal one, see Fig.~\ref{fig:witness2}. However, it is proved that the problem of accurately finding such an optimal witness is NP-hard \cite{ben1998robust}. Thus, our problem is also intractable for the most general case.

\subsection{$\epsilon$-level optimal EW}
The key for the problem being intractable is that there is no efficient way to characterize an arbitrary entanglement witness. In the bipartite case, an operator is an witness if and only if
\begin{equation}\label{eq:}
\mathrm{Tr}[\sigma_{AB}W]\ge0
\end{equation}
for any separable state $\sigma_{AB}$. As $\sigma_{AB}$ can always be decomposed as a convex combination of separable states as $\ket{\psi}_A\ket{\psi}_B$, the condition can be equivalently expressed as
\begin{equation}\label{eq:witnessconstraints}
\bra{\psi}_A\bra{\psi}_BW\ket{\psi}_A\ket{\psi}_B\ge0,
\end{equation}
for any pure states $\ket{\psi}_A$ and $\ket{\psi}_B$. The constraints for a witness $W$ are very difficult to describe in the most general case, which makes our problem hard.

While, this problem can be resolved if we allow certain failure errors. A Hermitian operator $W_\epsilon$ is defined as an $\epsilon$-level entanglement witness \cite{Brandao04}, when
\begin{equation}\label{eq:epsilonwitness}
\mathrm{Prob}\left\{\mathrm{Tr}[\sigma W_\epsilon]<0|\sigma \in S\right\} \le \epsilon,
\end{equation}
where $S$ is the set of separable states. That is, the operator $W_\epsilon$ has a probability less than $\epsilon$ to detect a randomly selected separable quantum state to be entangled. We can thus regard this $\epsilon$ as a failure error probability. It is shown that the $\epsilon$-level optimal EW can be found efficiently for any given entangled state $\rho$ \cite{Brandao04}. In particular, constrained on $\mathrm{Tr}[W_\epsilon] = 1$ and $W_\epsilon$ to be an $\epsilon$-level EW, one can run a semi-definite programming (SDP) to minimize $\mathrm{Tr}[W_\epsilon\rho]$.

\subsection{Solution}

Following the method proposed in Ref.~\cite{Brandao04}, we can solve the minimization problem given in Eq.~\eqref{eq:Jqminimum} by allowing a certain failure probability $\epsilon$. First, we relax the constraint given in Eq.~\eqref{eq:constraintsss}. Instead of requiring being non-negative for all separable states, we randomly generate $N$ separable states $\{\ket{\psi}_A^i\ket{\psi}_B^i\}$ and require that
\begin{equation}\label{eq:witnessconstraints2}
\sum_{x,y}\beta^{x,y}\langle\omega_x^{\mathrm{T}}\otimes \tau_y^{\mathrm{T}}\rangle^i\ge0, \forall i\in\{1,2,\dots,N\},
\end{equation}
where $\langle\omega_x^{\mathrm{T}}\otimes \tau_y^{\mathrm{T}}\rangle^i = \bra{\psi}_A^i\bra{\psi}_B^i \omega_x^{\mathrm{T}}\otimes \tau_y^{\mathrm{T}}\ket{\psi}_A^i\ket{\psi}_B^i$. Then the problem can be expressed as

\begin{flushleft}
\emph{Problem ($\epsilon$-level): given a probability distribution $p(1,1|\omega_x,\tau_y)$, minimize
\begin{equation}\label{eq:minimum}
J(\beta^{x,y}) = \sum_{x,y} \beta^{x,y}{p}(1,1|\omega_x,\tau_y)
\end{equation}
over all $\beta^{x,y}$ satisfying
\begin{equation}\label{eq:witnessconstraints3}
\sum_{x,y}\beta^{x,y}\langle\omega_x^{\mathrm{T}}\otimes \tau_y^{\mathrm{T}}\rangle^i\ge0, \forall i\in\{1,2,\dots,N\},
\end{equation}
for $N$ randomly generated separable states $\{\ket{\psi}_A^i\ket{\psi}_B^i\}$ and
\begin{equation}\label{eq:}
\sum_{x,y} \beta^{x,y}\mathrm{Tr}\left[\omega_x^{\mathrm{T}} \otimes \tau_y^{\mathrm{T}}\right] = 1.
\end{equation}
}
\end{flushleft}

This problem can be converted to an SDP solvable problem when we re-express the inequality of numbers in Eq.~\eqref{eq:witnessconstraints3} by an inequality of matrices. To do so, we only need to notice that Eq.~\eqref{eq:witnessconstraints} is equivalent to require that
\begin{equation}\label{eq:witnessconstraints2}
W_B = \bra{\psi}_AW_\epsilon\ket{\psi}_A\ge0, \forall \ket{\psi}_A,
\end{equation}
where $W_B\ge0$ indicates that $W_B$ has non-negative eigenvalues.
Therefore, we only need to generate $N$ states $\ket{\psi}_A^i$, for $i = 1, 2, \dots, N$, and the problem is
\begin{flushleft}
\emph{Problem ($\epsilon$-level, SDP): given a probability distribution ${p}(1,1|\omega_x,\tau_y)$, minimize
\begin{equation}\label{eq:minimum}
J(\beta^{x,y}) = \sum_{x,y} \beta^{x,y}{p}(1,1|\omega_x,\tau_y)
\end{equation}
over all $\beta^{x,y}$ satisfying
\begin{equation}\label{eq:witnessconstraints4}
\sum_{x,y}\beta^{x,y}\bra{\psi}_A^i\omega_x^{\mathrm{T}}\ket{\psi}_A^i\tau_y^{\mathrm{T}}\ge0, \forall i\in\{1,2,\dots,N\},
\end{equation}
for $N$ randomly generated states $\{\ket{\psi}_A^i\}$ and
\begin{equation}\label{eq:}
\sum_{x,y} \beta^{x,y}\mathrm{Tr}\left[\omega_x^{\mathrm{T}} \otimes \tau_y^{\mathrm{T}}\right] = 1.
\end{equation}
}
\end{flushleft}
In practice, we can run an SDP to solve this problem. According to Ref.~\cite{calafiore2005uncertain}, to get the $\epsilon$-level witness with probability at least $1-\beta$, the number of random states $N$ should be at least $r/(\epsilon\beta) - 1$. Here $r = (d_Ad_B)(d_Ad_B+1)$ and $\beta$ can be understood as the failure probability of the minimization program. It is worth to remark that the problem can be similarly solved in the multipartite case.


\section{Example}\label{Sec:Example}
In this section, we show explicit examples about how the witness becomes non-optimal in the MDI scenario and how this problem can be resolved by running the optimizing program.

Suppose the to-be-witnessed state is a two-qubit Werner state \cite{Werner89}:
\begin{equation}\label{eq:werner}
\rho^v_{AB} = v\ket{\Psi^-}\bra{\Psi^-} + (1-v)I/4,
\end{equation}
where $\ket{\Psi^-} = 1/\sqrt{2}(\ket{01}-\ket{10})$ and $I$ is the identity matrix. The designed entanglement witness for the Werner states is
\begin{equation}\label{eq:}
W = \frac{1}{2}I - \ket{\Psi^-}\bra{\Psi^-}.
\end{equation}
As $\mathrm{Tr}[W\rho^v_{AB}] = (1-3v)/4$, $\rho^v_{AB}$ is entangled for $v>1/3$ and separable otherwise.

As shown in Ref.~\cite{Branciard13}, we can choose the input set by
\begin{equation}\label{eq:}
\omega_x = \sigma_x\frac{I+\vec{n}\cdot\vec{\sigma}}{2}\sigma_x, \tau_y = \sigma_y\frac{I+\vec{n}\cdot\vec{\sigma}}{2}\sigma_y, x,y = 0,\dots, 3
\end{equation}
where $\vec{n} = (1,1,1)/\sqrt{3}$, $\vec{\sigma} =
(\sigma_1,\sigma_2,\sigma_3)$ is the Pauli matrices, and $\sigma_0 = I$. According to Eq.~\eqref{eq: decomposition}, the witness can be decomposed on the basis of $\{\omega_x\otimes\tau_y\}$ with coefficient $\beta^{x,y}$ given by
\begin{equation}\label{eq:}
\beta^{x,y} =\left\{\begin{array}{cc}
               \frac{5}{8},& \mathrm{if} \, x = y, \\
               -\frac{1}{8},& \mathrm{if} \, x \ne y.
             \end{array}\right.
\end{equation}
And the MDIEW value is given by
\begin{equation}\label{eq:}
J = \frac{5}{8}\sum_{x=y}p(1,1|\omega_x,\tau_y) - \frac{3}{8}\sum_{x\ne y}p(1,1|\omega_x,\tau_y).
\end{equation}

In the ideal case, the probability distribution $p(1,1|\omega_x,\tau_y)$ is obtained by projecting onto maximally entangled states, that is,
\begin{equation}\label{eq:}
p(1,1|\omega_x,\tau_y) = \mathrm{Tr}[\left(M_A \otimes M_B\right)\times (\omega_x \otimes \rho_{AB} \otimes \tau_y)]
\end{equation}
where $M_A = \ket{\Phi_{AA}^+}\bra{\Phi_{AA}^+}$ and $M_B = \ket{\Phi_{BB}^+}\bra{\Phi_{BB}^+}$. While, in practice, there may exist imperfection in measurement. For instance, we consider that Alice's measurement is perfect while Bob's measurement is instead
\begin{equation}\label{eq:}
M_B' = \ket{\Phi_{BB}^-}\bra{\Phi_{BB}^-},
\end{equation}
where $\ket{\Phi_{BB}^-} = 1/\sqrt{2}(\ket{00} - \ket{11})$. In the case of quantum key distribution, projecting onto $\ket{\Phi_{BB}^-}$ can be regarded as a phase error.

\begin{figure}[bht]
\centering
\resizebox{8cm}{!}{\includegraphics[scale=1]{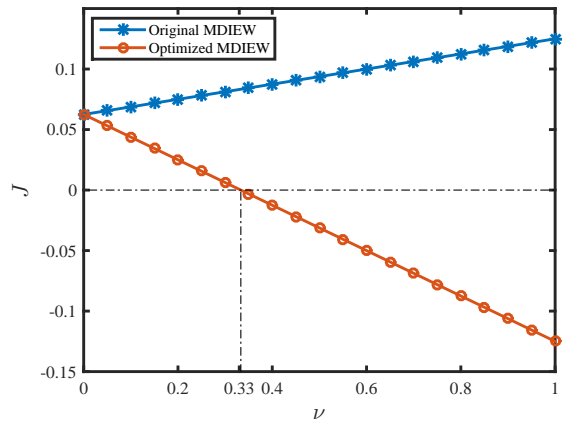}}
\caption{Simulation results of the original and optimized MDIEW protocol. The to be witness state is the two-qubit Werner state defined in Eq.~\eqref{eq:werner}. Here, we consider that Alice projects onto $\ket{\Phi_{AA}^+}$ and Bob projects onto $\ket{\Phi_{BB}^-}$. In this case, the original MDIEW cannot detect entanglement, while the optimized MDIEW protocol detects all entangle Werner states.}\label{fig:OMDIEW}
\end{figure}

As shown in Fig.~\ref{fig:OMDIEW}, we plot the MDIEW  and the optimized MDIEW results. For the original MDIEW result, as Bob's measurement is incorrect, no Werner state given in Eq.~\eqref{eq:werner} can be witnessed to be entangled. Although, by optimizing over all possible entanglement witness, we show that $\rho^v_{AB}$ is entangled as long as $v>1/3$. In this case, the optimized MDIEW can detect all entangled Werner states.

\section{Discussion}\label{Sec:Discussion}

In this work, we propose an optimized MDIEW scheme to solve the reliability and robust problem at the same time in entanglement detection, which maximally exploits the measurement data to investigate the entanglement property without trusting the measurement. By adopting $\epsilon$-level EW, we present an efficient way for the optimization procedure. As an explicit example, we show that the original MDIEW may not detect entanglement while our optimized MDIEW can. As our optimization can be regarded as a postprocessing of experiment data, our scheme can be easily applied in practice.

The optimization program finds the optimal $\epsilon$-level optimal EW $W_\epsilon$, which as its name indicates, has a probability less than $\epsilon$ to detect an separable state to be entangled. To get a smaller $\epsilon$, one can use a larger number $N$ of random states. In this case, the $\epsilon$ can be regarded as the statistical fluctuation which is inversely related to the number of trials $N$.
On the one hand, to efficiently get the optimal witness $W_\epsilon$, one has to introduce a nonzero failure error $\epsilon$; On the other hand, we can always add an extra term to the EW to eliminate $\epsilon$, i.e.,
\begin{equation}\label{}
  W = W_\epsilon + \alpha I,
\end{equation}
where $\alpha$ is chosen to be the minimum value such that $W$ is an entanglement witness. To efficiently find $\alpha$, one can make use of the technique similar to Ref.~\cite{Sperling13}, in which, EW can be systematically constructed.

\section*{Acknowledgments}
This work was supported by the 1000 Youth Fellowship program in China.

\appendix
\section{Proof of MDIEW}\label{App:proof}
Here, we review the properties of the MDIEW scheme and refer to Ref.~\cite{Branciard13} for further reference. First, when the measurement is to project onto the maximally entangled state, we show that
\begin{equation}\label{}
  J = \mathrm{Tr}[W\rho_{AB}]/\sqrt{d_Ad_B}.
\end{equation}
\begin{proof}
When Alice and Bob perform both projection measurement onto the maximally entangled state, the conditional probability distribution is
\begin{eqnarray}\label{eq: entanglementprobability}
	{p}(1,1|x,y) &=& \mathrm{Tr}[\left(\ket{\Phi_{AA}^+}\bra{\Phi_{AA}^+} \otimes \ket{\Phi_{BB}^+}\bra{\Phi_{BB}^+}\right)\times (\omega_x \otimes \rho_{AB} \otimes \tau_y)]\\
	&=& \mathrm{Tr}[(\omega_x^{\mathrm{T}} \otimes \tau_y^{\mathrm{T}}) \rho_{AB}]/\sqrt{d_A d_B}.
\end{eqnarray}
In this case, the MDIEW value $J$ is
\begin{eqnarray}\label{eq: entanglementMDIEW}
		J &= & \sum_{x,y} \beta^{x,y} \mathrm{Tr}[(\omega_x^{\mathrm{T}} \otimes \tau_y^{\mathrm{T}})\rho_{AB}]/\sqrt{d_A d_B}\\
			&= & \mathrm{Tr}[W\rho_{AB}] / \sqrt{d_A d_B}.
\end{eqnarray}
\end{proof}

Secondly, we show that for arbitrary measurement, the MDIEW value will be non-negative for any separable state $\sigma$.
\begin{proof}
Suppose Alice and Bob are asked to witness a separable state $\sigma_{AB} = \sum_i {p}_i \sigma_A^i \otimes \sigma_B^i$, where ${p}_i \ge 0, \sum_i {p}_i = 1$ and the measurements are general POVM elements $M_A, M_B$, respectively.
Consequently, we can represent the conditional probability distribution as following
\begin{eqnarray}\label{eq: separableprobability}
	{p}(1,1|x,y) &=& \mathrm{Tr}[(M_A \otimes M_B)(\omega_x \otimes \sigma_{AB} \otimes \tau_y)]\\
					   &=& \sum_i {p}_i \mathrm{Tr}[(A_1^i \otimes B_1^i)(\omega_x \otimes \tau_y)],
\end{eqnarray}
where $A_1^i = \mathrm{Tr}_A[M_A (I \otimes \sigma_A)]$ and $B_1^i = \mathrm{Tr}_B[M_B (I\otimes \sigma_B)]$. Therefore, the MDIEW value $J$ is
\begin{eqnarray}\label{eq: separableMDIEW}
	J &=& \sum_{x,y} \beta^{x,y} \sum_i {p}_i \mathrm{Tr}[(A_1^i \otimes B_1^i)(\omega_x \otimes \tau_y)]\\
		&=&  \mathrm{Tr}\left[\left(\sum_i {p}_iA_1^i \otimes B_1^i\right)W^{\mathrm{T}}\right]\\
&=&  \mathrm{Tr}\left[\left(\sum_i {p}_iA_1^i \otimes B_1^i\right)^{\mathrm{T}}W\right].
\end{eqnarray}
As $W$ is an EW and $\left(\sum_i {p}_iA_1^i \otimes B_1^i\right)^{\mathrm{T}}$ is a separable state, we can see that $J\ge 0$ for arbitrary measurement and separable state $\sigma_{AB}$.
\end{proof}

\section*{References}
 \bibliographystyle{iopart-num}

\bibliography{bibOMDIEW}

\providecommand{\newblock}{}
\begin{thebibliography}{10}
\expandafter\ifx\csname url\endcsname\relax
  \def\url#1{{\tt #1}}\fi
\expandafter\ifx\csname urlprefix\endcsname\relax\def\urlprefix{URL }\fi
\providecommand{\eprint}[2][]{\url{#2}}

\bibitem{Einstein35}
Einstein A, Podolsky B and Rosen N 1935 {\em Phys. Rev.\/} {\bf 47}(10)
  777--780 \urlprefix\url{http://link.aps.org/doi/10.1103/PhysRev.47.777}

\bibitem{bell}
Bell J~S 1987 {\em On the Einstein-Podolsky-Rosen Paradox. Physics 1, 195--200
  (1964)\/} Speakable and Unspeakable in Quantum Mechanics (Cambridge
  University Press)

\bibitem{Horodecki09}
Horodecki R, Horodecki P, Horodecki M and Horodecki K 2009 {\em Rev. Mod.
  Phys.\/} {\bf 81}(2) 865--942
  \urlprefix\url{http://link.aps.org/doi/10.1103/RevModPhys.81.865}

\bibitem{bb84}
Bennett C~H and Brassard G 1984 {Quantum Cryptography: Public Key Distribution
  and Coin Tossing} {\em Proceedings of the IEEE International Conference on
  Computers, Systems and Signal Processing\/} (New York: IEEE Press) pp
  175--179

\bibitem{Ekert91}
Ekert A~K 1991 {\em Phys. Rev. Lett.\/} {\bf 67}(6) 661--663
  \urlprefix\url{http://link.aps.org/doi/10.1103/PhysRevLett.67.661}

\bibitem{Lutkenhaus04}
Curty M, Lewenstein M and L{\"u}tkenhaus N 2004 {\em Phys. Rev. Lett.\/} {\bf
  92} 217903

\bibitem{tura2014detecting}
Tura J, Augusiak R, Sainz A, V{\'e}rtesi T, Lewenstein M and Ac{\'\i}n A 2014
  {\em Science\/} {\bf 344} 1256--1258

\bibitem{guhne2009}
G{\"u}hne O and T{\'o}th G 2009 {\em Physics Reports\/} {\bf 474} 1--75

\bibitem{Yuan14}
Xu P, Yuan X, Chen L~K, Lu H, Yao X~C, Ma X, Chen Y~A and Pan J~W 2014 {\em
  Phys. Rev. Lett.\/} {\bf 112}(14) 140506
  \urlprefix\url{http://link.aps.org/doi/10.1103/PhysRevLett.112.140506}

\bibitem{Lo2012MDI}
Lo H~K, Curty M and Qi B 2012 {\em Phys. Rev. Lett.\/} {\bf 108}(13) 130503
  \urlprefix\url{http://link.aps.org/doi/10.1103/PhysRevLett.108.130503}

\bibitem{Biham:1996:Quantum}
Biham E, Huttner B and Mor T 1996 {\em Phys. Rev. A\/} {\bf 54}(4) 2651--2658
  \urlprefix\url{http://link.aps.org/doi/10.1103/PhysRevA.54.2651}

\bibitem{Inamori:TimeReverseEPR:2002}
Inamori H 2002 {\em Algorithmica\/} {\bf 34} 340 ISSN 0178-4617
  \urlprefix\url{http://dx.doi.org/10.1007/s00453-002-0983-4}

\bibitem{Stefano:MDIQKD:2012}
Braunstein S~L and Pirandola S 2012 {\em Phys. Rev. Lett.\/} {\bf 108}(13)
  130502 \urlprefix\url{http://link.aps.org/doi/10.1103/PhysRevLett.108.130502}

\bibitem{Branciard13}
Branciard C, Rosset D, Liang Y~C and Gisin N 2013 {\em Phys. Rev. Lett.\/} {\bf
  110}(6) 060405
  \urlprefix\url{http://link.aps.org/doi/10.1103/PhysRevLett.110.060405}

\bibitem{Buscemi12}
Buscemi F 2012 {\em Phys. Rev. Lett.\/} {\bf 108}(20) 200401
  \urlprefix\url{http://link.aps.org/doi/10.1103/PhysRevLett.108.200401}

\bibitem{boyd2004convex}
Boyd S and Vandenberghe L 2004 {\em Convex optimization\/} (Cambridge
  university press)

\bibitem{ben1998robust}
Ben-Tal A and Nemirovski A 1998 {\em Mathematics of Operations Research\/} {\bf
  23} 769--805

\bibitem{Werner89}
Werner R~F 1989 {\em Phys. Rev. A\/} {\bf 40}(8) 4277--4281
  \urlprefix\url{http://link.aps.org/doi/10.1103/PhysRevA.40.4277}

\bibitem{Barrett02}
Barrett J 2002 {\em Phys. Rev. A\/} {\bf 65}(4) 042302
  \urlprefix\url{http://link.aps.org/doi/10.1103/PhysRevA.65.042302}

\bibitem{Massar03}
Massar S and Pironio S 2003 {\em Phys. Rev. A\/} {\bf 68}(6) 062109
  \urlprefix\url{http://link.aps.org/doi/10.1103/PhysRevA.68.062109}

\bibitem{Wilms08}
Wilms J, Disser Y, Alber G and Percival I~C 2008 {\em Phys. Rev. A\/} {\bf
  78}(3) 032116
  \urlprefix\url{http://link.aps.org/doi/10.1103/PhysRevA.78.032116}

\bibitem{Brandao04}
Brand\~ao F~G~S~L and Vianna R~O 2004 {\em Phys. Rev. Lett.\/} {\bf 93}(22)
  220503 \urlprefix\url{http://link.aps.org/doi/10.1103/PhysRevLett.93.220503}

\bibitem{calafiore2005uncertain}
Calafiore G and Campi M~C 2005 {\em Mathematical Programming\/} {\bf 102}
  25--46

\bibitem{Sperling13}
Sperling J and Vogel W 2013 {\em Phys. Rev. Lett.\/} {\bf 111}(11) 110503
  \urlprefix\url{http://link.aps.org/doi/10.1103/PhysRevLett.111.110503}

\end{thebibliography}

\end{document}